\documentclass[]{spie}  %>>> use for US letter paper
\usepackage[]{graphicx}

\title{Using ACIS on the Chandra X-ray Observatory as a particle radiation monitor II} 

\author{C. E. Grant\supit{a}, P. G. Ford\supit{a}, M. W. Bautz\supit{a} and S. L. O'Dell\supit{b}
\skiplinehalf
\supit{a}Kavli Institute for Astrophysics and Space Research, Massachusetts Institute of Technology, Cambridge, Massachusetts, USA; \\
\supit{b}NASA Marshall Space Flight Center, Huntsville, Alabama, USA
}

\authorinfo{Further author information: (Send correspondence to C.E.G.)\\C.E.G.: E-mail: cgrant@space.mit.edu}

\begin{document} 
  \maketitle 

\begin{abstract}
The Advanced CCD Imaging Spectrometer is an instrument on the Chandra X-ray Observatory. CCDs are vulnerable to radiation damage, particularly by soft protons in the radiation belts and solar storms. The Chandra team has implemented procedures to protect ACIS during high-radiation events including autonomous protection triggered by an on-board radiation monitor. Elevated temperatures have reduced the effectiveness of the on-board monitor. The ACIS team has developed an algorithm which uses data from the CCDs themselves to detect periods of high radiation and a flight software patch to apply this algorithm is currently active on-board the instrument. In this paper, we explore the ACIS response to particle radiation through comparisons to a number of external measures of the radiation environment. We hope to better understand the efficiency of the algorithm as a function of the flux and spectrum of the particles and the time-profile of the radiation event.
\end{abstract}


\section{INTRODUCTION}
\label{sec:intro}

The Chandra X-ray Observatory, the third of NASA's great observatories in space, was launched just past midnight on July 23, 1999, aboard the space shuttle {\it Columbia}\cite{cha2}.  After a series of orbital maneuvers, Chandra reached its operational orbit, with initial 10,000-km perigee altitude, 140,000-km apogee altitude, and 28.5$^\circ$ inclination.  In this evolving high elliptical orbit, Chandra transits a wide range of particle environments, from the radiation belts at closest approach through the magnetosphere and magnetopause and past the bow shock into the solar wind.

The Advanced CCD Imaging Spectrometer (ACIS), one of two focal plane science instruments on Chandra, utilizes frame-transfer charge-coupled devices (CCDs) of two types, front- and back-illuminated (FI and BI)\cite{acis}.  Soon after launch it was discovered that the FI CCDs had suffered radiation damage from exposure to soft protons scattered off the Observatory's grazing-incidence optics during passages through the Earth's radiation belts\cite{gyp00}.  Since mid-September 1999, ACIS has been protected during radiation belt passages and there is an ongoing effort to prevent further damage and to develop hardware and software strategies to mitigate the effects of radiation damage on data analysis\cite{odell}.

Our primary measure of radiation damage on the CCDs is charge transfer inefficiency (CTI).  The eight front-illuminated CCDs had essentially no CTI before launch, but are strongly sensitive to radiation damage from low energy protons ($\sim$100~keV) which preferentially create traps in the buried transfer channel.  The framestore covers are thick enough to stop this radiation, so the initial damage was limited to the imaging area of the FI CCDs.  Radiation damage from low-energy protons is now minimized by moving the ACIS detector away from the aimpoint of the observatory during passages through the Earth's particle belts and during solar storms.  Continuing exposure to both low and high energy particles over the lifetime of the mission slowly degrades the CTI further.\cite{odell,ctitrend}  The two back-illuminated CCDs (ACIS-S1,S3) suffered damage during the manufacturing process and exhibit CTI in both the imaging and framestore areas and the serial transfer array.  However, owing to their much deeper charge-transfer channel, BI CCDs are insensitive to damage by the low-energy ions that damage FI devices.

Since early in the Chandra mission, procedures have been implemented that protect the focal plane instruments during times of high radiation. \cite{odell} ACIS is translated out of the focal plane, providing protection against soft protons, and is powered off.  Three types of procedures are in place; planned protection during radiation-belt transits, autonomous protection triggered by the on-board radiation monitor, and manual intervention based upon assessment of space-weather conditions.  The Chandra weekly command load includes automatic scheduled safing of the focal-plane instruments during radiation belt passages.  The timing of radiation belt ingress and egress are determined using the standard AP8/AE8 environment with a small additional pad time to protect against temporal variations.  Solar storms are detected either by the on-board radiation monitor or by ground operations monitoring of various space weather measures, such as from NASA's Advanced Composition Explorer (ACE)\cite{ace}, the NOAA Geostationary Operational Environmental Satellite system (GOES), and the planetary index Kp.  The on-board radiation monitor cannot detect protons at hundreds of keV, which are the most damaging to ACIS, so on-board protection is supplemented by other measures of the radiation environment. 

The Electron, Proton, Helium INstrument (EPHIN) is a particle detector on-board the Chandra spacecraft used to monitor the local particle radiation environment.  It is sensitive to electrons in the energy range 150~keV--50~MeV and protons from 5--49~MeV.  Chandra-EPHIN is very similar to the EPHIN detector onboard SOHO.\cite{ephin}  Until December 2008, EPHIN rates in three channels were monitored by the spacecraft computer which can command radiation shutdowns during solar storms.  The monitored channels were P4, sensitive to protons with 5.0--8.3~MeV, P41, sensitive to protons with 41--53~MeV and E1300, sensitive to electrons with 2.64--6.18~MeV.  As the spacecraft insulation has aged and degraded, thermal control of some subsystems, including EPHIN, has become more difficult.  Elevated EPHIN temperatures cause anomalous noise, which in some EPHIN channels can be significant and occasionally dominate the signal.\cite{odell}  To prevent against false triggers due to EPHIN noise, EPHIN was reconfigured in December 2008 into a mode that does not distinguish particle species.  At that point, the spacecraft computer began monitoring only two EPHIN channels, one of which was subsequently deleted due to increasing noise.  The remaining monitored EPHIN channel, E1300$^\prime$, is basically the union of the original E1300 and P41 channels; hence, it is sensitive to high-energy (41--53 MeV) protons.  Concern about the effectiveness of EPHIN going into the future motivates looking for other on-board measures of the radiation environment.  Consequently, the spacecraft computer now monitors the anti-coincidence shield rate of the other Chandra focal-plane instrument, the High Resolution Camera (HRC)\cite{odell}.  As the HRC shield responds to only penetrating charged particles (protons $>$ 30~MeV), it monitors approximately the same proton energies as does the E1300$^\prime$ channel.  A second measure, examined here, is using ACIS itself.

We first addressed the potential use of ACIS as its own particle monitor in Ref.~\citenum{radmon1} (Paper I).  As the primary purpose of ACIS is always to collect astrophysical data, the particle monitoring processes need to be secondary in using the available on-board resources and need to be flexible in dealing with the numerous observing modes and X-ray source types.  Ideally the scientific user should be completely unaware of the simultaneous radiation monitoring during his/her observation.  The monitor needs to detect as many instances of enhanced radiation environment as possible, without unnecessarily interrupting observations due to bright X-ray sources or hardware anomalies.

Paper I was an initial exploration into the efficacy of a particular proxy for the radiation environment---the threshold crossing rate.  ACIS performs X-ray event finding and recognition tasks on-board in real time.  Each CCD frame is examined to find pixels with pulseheights above a pre-determined threshold value.  These pixels are considered candidates to be X-ray events and are subjected to further processing.  The number of pixels above the event detection threshold in a CCD frame, or threshold crossing rate, is calculated and saved as part of the standard event processing on-board and is sensitive to both the X-ray and particle intensity.  We have developed an ACIS flight software patch that keeps track of the mean threshold crossing rate (pixels/second/row) for both FI and BI CCDs and signals the Chandra On-Board Computer (OBC) when this rate exceeds a predetermined value and is increasing.  Ref.~\citenum{pgfradmon} from this conference describes in more detail the algorithm, the software patch, and determination of the optimal parameters for the patch.

The flight software patch was installed on ACIS in November 2011 and updated with more optimal parameters in April 2012.  The patch has operated as expected with no impact on regular science operations and has correctly indicated an enhanced radiation environment on two occasions.  The Chandra OBC was patched in May 2012 and will now respond to any ACIS radiation triggers with the standard radiation protection procedures.  To better match the changing quiescent particle background which is anti-correlated with the solar cycle, the parameters of the patch will be re-evaluated a few times a year and updated as necessary.

In this paper, we explore the response of the ACIS radiation monitor to the Chandra particle environment.  In particular, we examine the historical data to better understand the nature of the particle events that would and would not have triggered the ACIS radiation monitor.

\section{ACIS DATA}

The ACIS flight software patch uses the event threshold crossing rate to monitor the radiation environment.  In particular, it keeps track of the threshold crossings per second per row, which is, for the most part, independent of the instrument observing mode.  The ACIS focal plane has two types of CCD, front-illuminated (FI) and back-illuminated (BI), which have different response to particle events.  To reduce the noise in the threshold rates, the radiation monitor patch examines the mean of the active FI CCDs and the mean of the active BI CCDs, in three minute time bins.  Depending on the observing mode, there may be as many as two BI CCDs or six FI CCDs, or as few as a single CCD of a given type.  In cases with fewer active CCDs, the mean threshold crossing rate will be noisier, potentially impacting the efficacy of the software patch.

\begin{figure}
\begin{center}
\includegraphics[height=6in,angle=270]{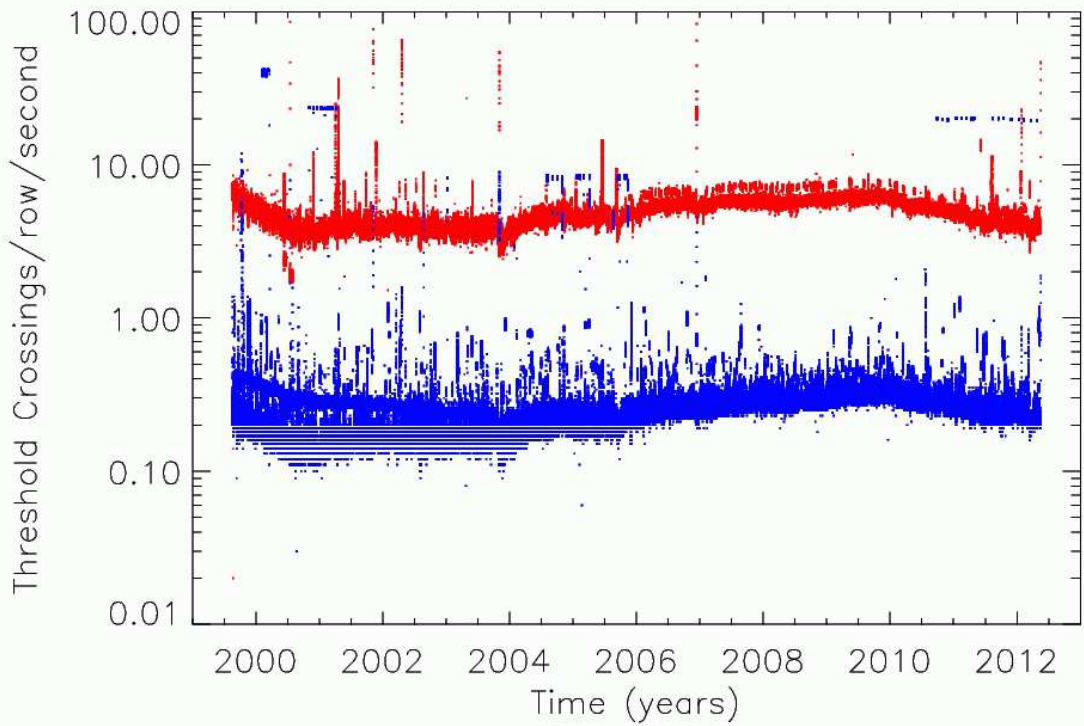}
\end{center}
\caption{ACIS threshold crossing rate as a function of time over the entire history of Chandra.  The (red) upper points denote the mean of the FI CCDs; the (blue) lower points, the BI CCDs.  The features are explained in the text.}
\label{fig:threshplot}
\end{figure}

The number of threshold crossings in each frame for each CCD is telemetered as part of the standard exposure record.  We can, therefore, examine the entire nearly thirteen year history of threshold crossing rates from long before the ACIS radiation monitor patch was even contemplated.  Figure~\ref{fig:threshplot} shows the threshold crossing rate as a function of time from August 1999 through May 2012.  The data from each type of CCD, FI and BI, have been averaged and put into three minute time bins, as is done by the software patch.  This threshold crossing rate includes contributions both from X-ray sources in the field and from the particle background.

A number of features can be seen in Figure~\ref{fig:threshplot}.  The threshold crossing rate of the FI CCDs is about an order of magnitude higher than that of the BI CCDs.  This is due to the structural differences of the two types of CCDs.  Particle interactions in the FI CCDs produce large blooms due to the thicker active and field-free regions.  The BI CCDs have a thinner active region and no field-free region, so the particle events occupy much smaller areas and produce fewer pixels above the event threshold.  There is a long timescale structure in the quiescent rates with lower threshold crossing rates during 2001--2004 and higher rates in 2008--2010, due to the 11-year solar cycle.  During solar maximum the sun's magnetic field provides extra shielding against cosmic rays which depresses the threshold crossing rate.

Times of enhanced threshold crossing rates are also seen in Figure~\ref{fig:threshplot}.  Many of the vertical enhancements correspond to real increases in the particle environment associated with solar activity.  The horizontal features seen in the BI CCD data are due to repeated observations of a specific bright X-ray source, the Crab Nebula.  A few cases of higher threshold crossing rates are due to instrument anomalies.

\section{EFFICIENCY OF RADIATION TRIGGERS}

We can apply the algorithm used by the ACIS radiation monitor to the accumulated history of ACIS threshold crossing data to find instances in which the algorithm indicates an elevated radiation environment.  By comparing to other external measures of the particle levels, we can also confirm that the radiation trigger is real.  Conversely, we have a record of times in which the on-board EPHIN particle detector triggered a radiation shutdown, which we can compare to the ACIS-based triggers.  Ideally, the ACIS radiation monitor would essentially duplicate the historical performance of EPHIN, but that cannot be the case due to the constraints of ACIS data taking.

One difficulty of using actual ACIS data is that ACIS, by the design of the radiation protection plan, is turned off when the radiation environment is believed to be high.  The ACIS data is truncated, limiting our ability to find times of high radiation using ACIS and to correlate with other radiation measures.  Testing using the historical data can only yield a lower limit on the number of times an ACIS-based radiation monitor would have triggered, as we have no direct information as to what the ACIS threshold crossing rate would be doing after a radiation shutdown.  In the absence of EPHIN, the ACIS radiation monitor would likely have more triggers than are found here.

A strength of using the real telemetry in testing the algorithm is that it reflects the true heterogeneity of ACIS observing modes and X-ray source types.  Any automated radiation monitor must be robust against false triggers due to high X-ray source rates or unusual ACIS science modes.  In developing the software algorithm, it became apparent immediately that bright X-ray sources could potentially cause false triggers.  For this reason, the algorithm requires that the threshold crossing rates are both high and monotonically increasing for five consecutive time bins.  After this addition, there were no false triggers due to bright X-ray sources.  This does, however, limit the ability of the ACIS radiation monitor to detect an ongoing radiation event if the initial increase was not observed or if the initial increase was slow or noisy.

By definition, an ACIS radiation monitor is blind while ACIS is not taking data.  This includes the initial period while the CCD bias level is being calculated, which can take up to twenty minutes, but also during radiation belt passages when the instrument is powered off for many hours, and any time the other focal plane instrument is observing the sky.  If the initial increase in particle rates is entirely encompassed in one of these time periods, the ACIS radiation monitor algorithm is much less likely to produce a trigger.

\begin{figure}
\includegraphics[height=8.1in]{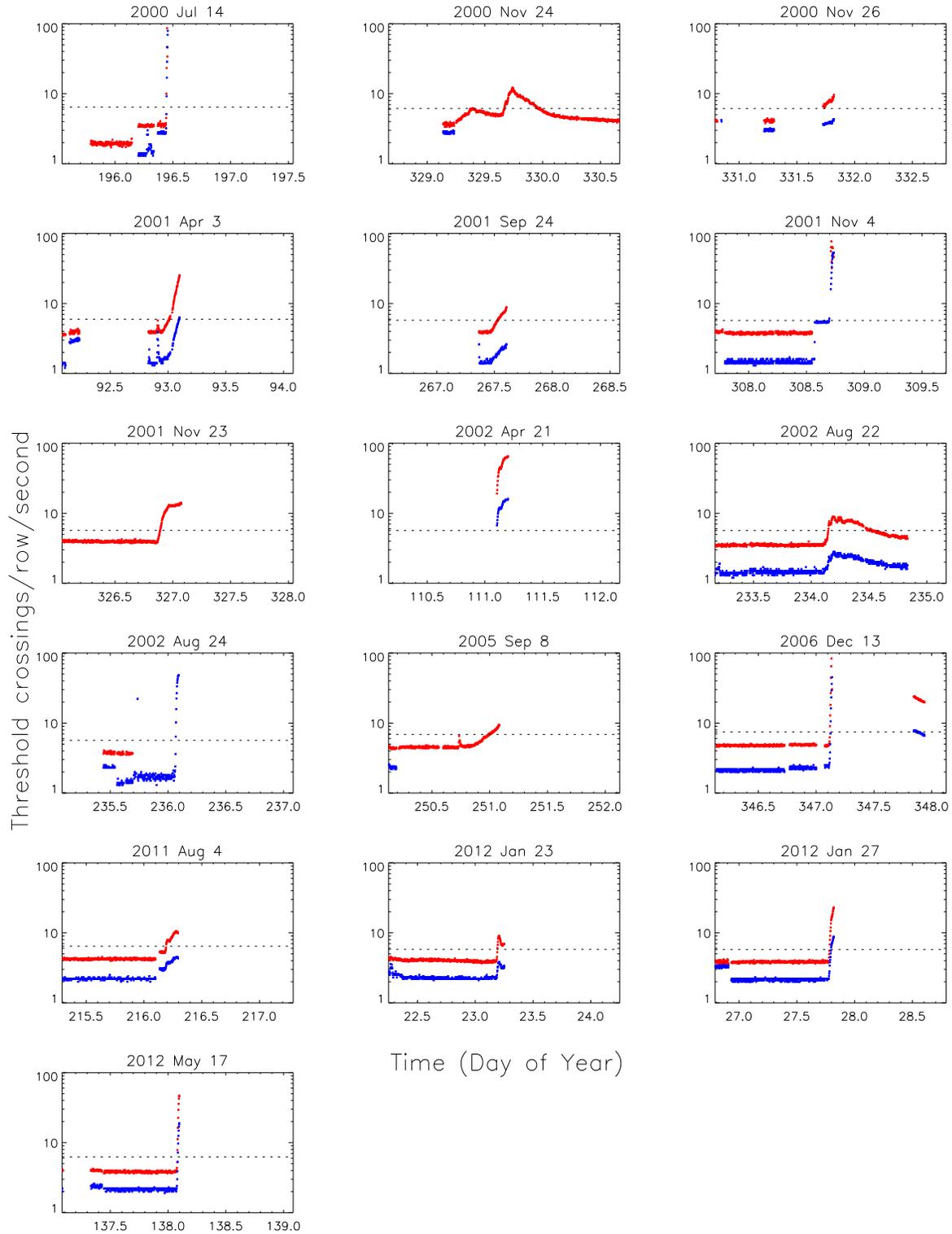}
\caption{The threshold crossing rates in a two day window around sixteen radiation triggers found by the ACIS radiation monitor algorithm.  The upper data (red) are FI CCD rates; the lower data (blue), BI CCD rates multiplied by ten.  The horizontal dotted line is the trigger threshold level for the FI CCDs.  The rates must be above this level and monotonically increasing for five data points to trigger the algorithm.  Gaps occur when the instrument is not taking data.}
\label{fig:triggerplots}
\end{figure}

The ACIS radiation monitor algorithm with the optimal parameters that are currently on-board the spacecraft has been applied to the entire history of ACIS data.  Sixteen triggers were found in which the threshold crossing rates are high and increasing.  All sixteen are during periods with enhanced radiation environment as measured by other instruments.  Figure~\ref{fig:triggerplots} shows the threshold crossing rates in the two days surrounding each of the sixteen ACIS radiation triggers.  In all but two cases, the ACIS trigger occurs shortly before an on-board EPHIN trigger which shut down the instrument, truncating the ACIS data.  In the remaining two cases the EPHIN rates do indicate an enhanced radiation environment, but did not quite meet the criteria for a shutdown.

Four of the ACIS radiation triggers occurred while the instrument was out of the focal plane and observing its radioactive calibration source.  Observations of the on-board calibration source are routinely scheduled immediately before and after perigee passages.  Current spacecraft operating procedure is to disable the autonomous radiation shutdown commands once the instruments have been moved out of the focal plane, so for a brief period of time ACIS is observing its calibration source without autonomous radiation protection.  As the calibration source position is protected from soft particles traveling through the optics, this is considered acceptable, however it does mean that an ACIS radiation trigger during this time will be ignored by the spacecraft OBC.

\begin{figure}
\includegraphics[height=7.3in]{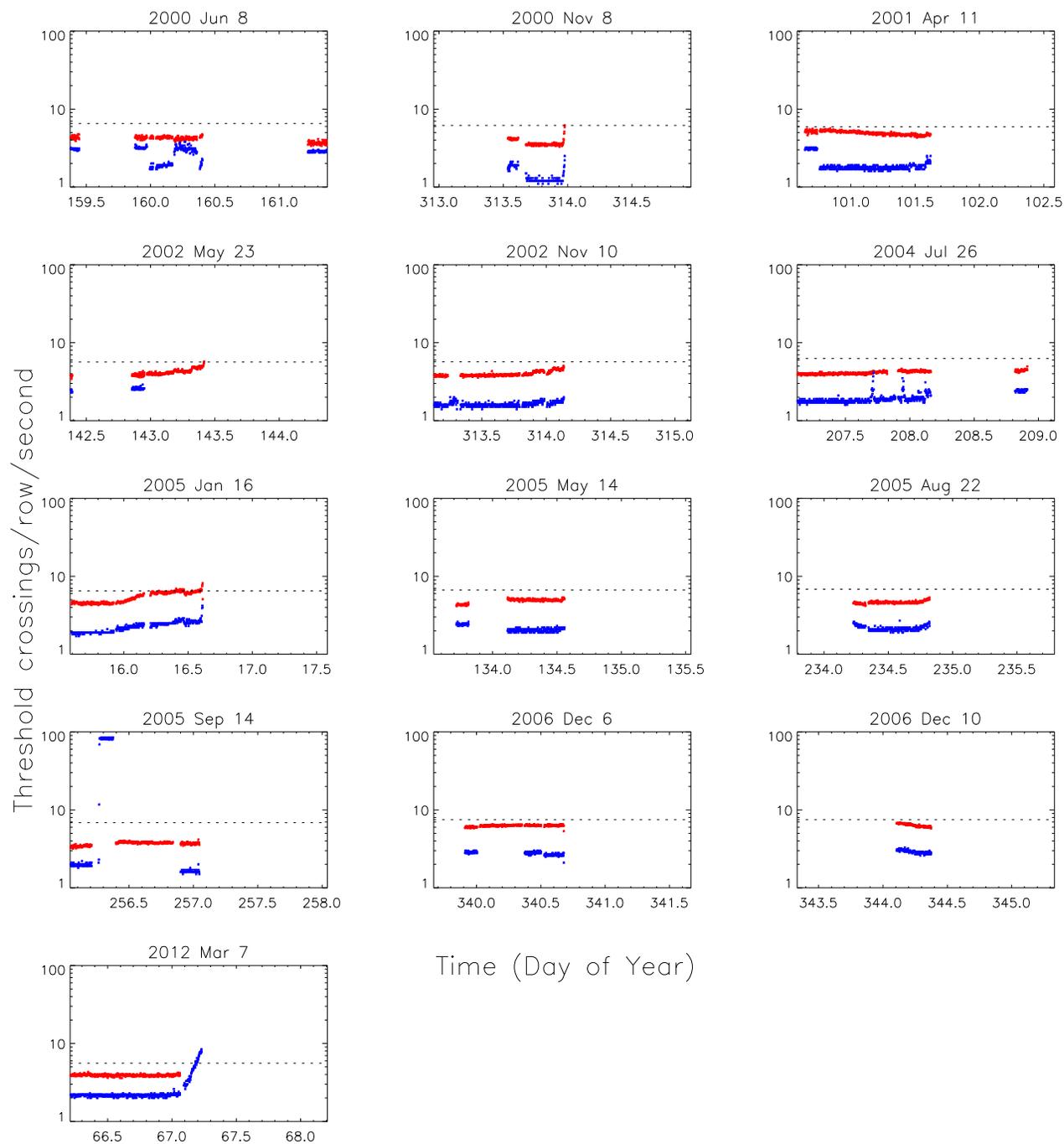}
\caption{The threshold crossing rates in a two day window around the thirteen radiation shutdowns that did not trigger the ACIS radiation monitor algorithm.  The upper data (red) are FI CCD rates; the lower data (blue), BI CCD rates multiplied by ten.  The horizontal dotted line indicates the trigger threshold level for the FI CCDs.  The rates must be above this level and monotonically increasing for five data points to trigger the algorithm.  Gaps occur when the instrument is not taking data.}
\label{fig:notriggerplots}
\end{figure}

There are an additional thirteen instances when the on-board EPHIN initiated a shutdown from a solar storm while ACIS was in the focal plane taking data, in which the ACIS radiation monitor algorithm does not trigger.  These are shown in Figure~\ref{fig:notriggerplots}.  A few of these (2000 Nov 8 for example) do show enhanced threshold crossing rates that, if not interrupted by the EPHIN shutdown, might have continued on to trigger the ACIS radiation monitor algorithm.  An interesting case is 7 Mar 2012.  Only one BI CCD was taking data, so while the threshold crossing rates were high enough, the data is noisy and not monotonically increasing.  Many of these shutdowns that did not trigger the ACIS algorithm, however, show slower increases, if any at all.

There are additional Chandra radiation shutdowns that we are not discussing here.  These include cases where there is no concurrent ACIS data, but also cases in which the shutdown was commanded from the ground based upon observations of a potential high radiation environment from other instruments, primarily ACE and GOES.  These manual shutdowns occur not at the precise moment the high radiation is expected to occur but at the next available spacecraft contact after the prediction of high radiation has been made.  Therefore, the initial increase in particle rates will not necessarily be observed by ACIS and so we have decided not to include these in the count of ``missed" radiation shutdowns.  In fact, one of the manual shutdowns is one of the sixteen ACIS radiation triggers, in which the initial increase occurred during the calibration source observations immediately before radiation belt entry.  The shutdown was commanded hours later when Chandra was exiting the radiation belts.

In summary, there were ten instances where ACIS was in the focal plane taking data during an EPHIN radiation shutdown that would have triggered the ACIS radiation monitoring patch.  There were also two ACIS triggers due to high radiation environment without a corresponding EPHIN shutdown and four ACIS triggers while ACIS was out of the focal plane.  These numbers can be compared to thirteen instances of an EPHIN shutdown due to a solar storm when ACIS was taking data that did not trigger the ACIS radiation patch.  While ACIS may not be an ideal particle radiation detector, it certainly does appear to provide some protection, and should be an effective addition to the Chandra radiation protection program.

\section{CHARACTERISTICS OF RADIATION TRIGGERS}

We wish to better understand the characteristics of the solar storms that did trigger the ACIS radiation monitor, and those that did not.  To begin with, we can examine Figures~\ref{fig:triggerplots} and \ref{fig:notriggerplots}.  Most of the triggers show very sharp initial increases in the ACIS threshold crossings rate, jumping up by an order of magnitude in less than half an hour.  With a few exceptions, the non-triggering radiation shutdowns show no evidence of even the beginnings of such an increase.  In order to determine whether they might have looked more similar if the data had not been truncated, we need to examine other measures of the particle background.

\begin{figure}
\includegraphics[height=7in]{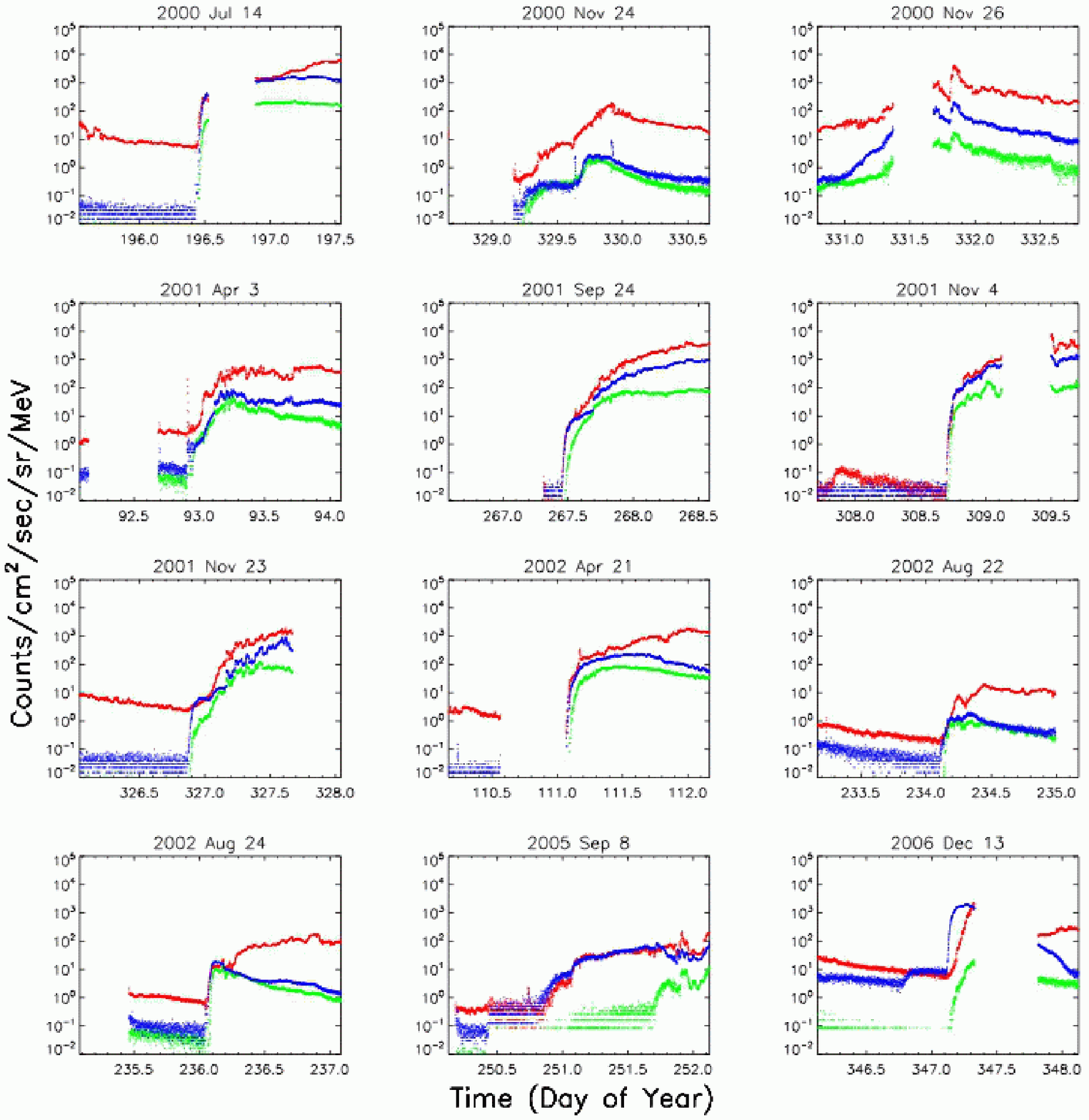}
%\vspace{4in}
%\special{psfile=triggerplots.ephin.test.ps angle=0 hscale=97 vscale=97 voffset=20 hoffset=0}
\caption{The EPHIN particle rates in a two day window around twelve radiation triggers found by the ACIS radiation monitor algorithm before 2008.  The (red) upper trace is P4 (5.0--8.3~MeV protons); the (blue) middle trace, E1300 (2.64--6.18~MeV electrons); and the (green) bottom trace, P41 (41--53~MeV protons).  Data during radiation belt passages are omitted for clarity.}
\label{fig:triggerephin}
\end{figure}

Figure~\ref{fig:triggerephin} shows the EPHIN particle rates during the same radiation shutdowns shown in Figure~\ref{fig:triggerplots}.  The three colors correspond to the three particle channels that were then used to monitor the radiation environment.  As EPHIN was reconfigured in 2008, and is currently using different channels, we are only showing the twelve shutdowns before this time for clarity.  While there are a range of radiation event time profiles, many of them do demonstrate a sharp increase of many orders of magnitude in particle rates over a short period of time.  In most cases, all three EPHIN channels are increasing similarly.

\begin{figure}
\includegraphics[height=7in]{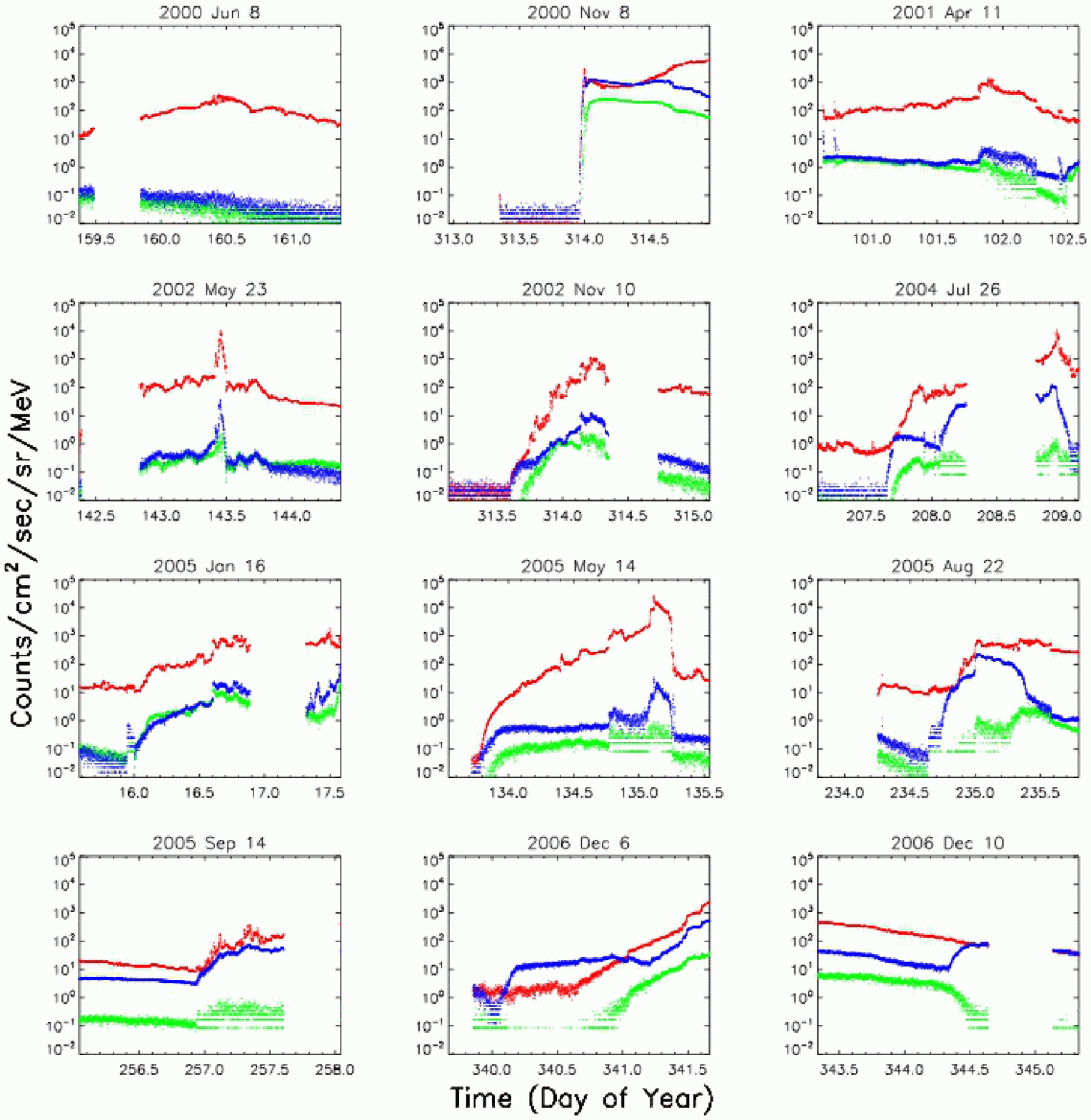}
%\vspace{7in}
%\special{psfile=nontriggerplots.ephin.eps angle=0 hscale=97 vscale=97 voffset=20 hoffset=0}
\caption{The EPHIN particle rates in a two day window around twelve radiation shutdowns before 2008 that did not trigger the ACIS radiation monitor algorithm.  The (red) upper trace is P4 (5.0--8.3~MeV protons); the (blue) middle trace, E1300 (2.64--6.18~MeV electrons); and the (green) bottom trace, P41 (41--53~MeV protons).  Data during radiation belt passages are omitted for clarity.}
\label{fig:notriggerephin}
\end{figure}

For comparison, Figure~\ref{fig:notriggerephin} shows the same EPHIN channels for the twelve radiation shutdowns before 2008 that did not trigger the ACIS radiation monitor algorithm.  These are much more heterogeneous than those in Figure~\ref{fig:triggerephin}.  With the exception of 2000 Nov 8, they exhibit much slower or noisier rises that would have difficulty meeting the monotonically increasing criteria.  The 2000 Nov 8 radiation shutdown is the one case in which it seems likely that an ACIS trigger would have occurred if the data hadn't been truncated.

Other characteristics of the EPHIN particle radiation during shutdowns, such as peak intensity or proton spectral hardness, do not seem to correlate well with the response of the ACIS radiation monitor.  There are examples of very high and more moderate intensity level in both the ACIS triggers and non-triggers, as well as harder and softer spectra.  It is possible that by examining data from other particle detectors, some further correlations might be found.  In particular, the EPHIN channels are not sensitive to protons less energetic than 5~MeV.  Preliminary examination of ACE data, which is sensitive to the low-energy ($\sim$100~keV) protons that damage the FI devices, does not seem to indicate any strong correlation with the ACIS radiation triggers either.

The clearest difference between the radiation shutdowns that triggered the ACIS radiation monitor and those that did not, seems to be the time profile of the particle intensity.  Events with a slower and noisier rise are less likely to trigger the ACIS radiation monitor, due to the requirement for triggers to have monotonically increasing count rates.  Other particle measures do not appear to be well correlated with whether or not a radiation event triggers the ACIS radiation monitor.  Luckily, the strongest need for autonomous radiation protection is from events that ramp up too quickly for ground intervention to be effective, so the ACIS radiation monitor is a useful addition to Chandra radiation protection measures.

\acknowledgments     %>>>> equivalent to \section*{ACKNOWLEDGMENTS}       
We would like to thank Paul Plucinsky, Scott Wolk, the Chandra Science Operations team, and the Chandra Project Science team for many years of fruitful collaboration and constant vigilance in managing ACIS radiation damage.  This work was supported by NASA contracts NAS 8-37716 and NAS 8-38252. 
 
\bibliography{article}   %>>>> bibliography data in report.bib
\bibliographystyle{spiebib}   %>>>> makes bibtex use spiebib.bst

\end{document}